\documentclass[a4paper]{article}
\usepackage{amsmath,amssymb}
\numberwithin{equation}{section}
\newfont{\wasy}{wasy10}

\newcommand{\dd}{{\mathbf d}}
\newcommand{\cN}{{\mathcal N}}
\newcommand{\cR}{{\mathcal R}}
\newcommand{\RR}{{\Bbb R}}
\newcommand{\scri}{{\mathcal J}}
\newcommand{\del}{\partial}
\newcommand{\eps}{\epsilon}
\newcommand{\epsb}{\bar\epsilon}
\newcommand{\nablab}{\overline{\nabla}}
\newcommand{\mb}{\overline{m}}
\newcommand{\mub}{\bar\mu}
\newcommand{\lambdab}{\overline\lambda}
\newcommand{\taub}{\bar\tau}
\newcommand{\sigmab}{\bar\sigma}
\newcommand{\gammab}{\bar\gamma}
\newcommand{\zetab}{\bar\zeta}
\newcommand{\omegab}{\overline{\omega}}

\newcommand{\thorn}{\mbox{\wasy\symbol{'151}}}

\title {On an integral formula on hypersurfaces in General Relativity}
\author{J\"org Frauendiener\\
Institut f\"ur Theoretische Astrophysik,\\
Universit\"at T\"ubingen,\\
Auf der Morgenstelle 10,\\
D-72076 T\"ubingen,\\
Germany}
\begin{document}
\maketitle

\begin{abstract}
We derive a general integral formula on an embedded hypersurface for
general relativistic space-times. Suppose the hypersurface is foliated
by two-dimensional compact ``sections'' $S_s$. Then the formula
relates the rate of change of the divergence of outgoing light rays
integrated over $S_s$ under change of section to geometric (convexity
and curvature) properties of $S_s$ and the energy-momentum content of
the space-time. We derive this formula using the
Sparling-Nester-Witten identity for spinor fields on the hypersurface
by appropriate choice of the spinor fields. We discuss several special
cases which have been discussed in the literature before, most notably
the Bondi mass loss formula.
\end{abstract}

\section{Introduction}
\label{sec:intro}

The purpose of this article is to derive and to discuss an integral
formula which can be obtained from the part of the Einstein equations
which is ``intrinsic'' to a hypersurface. This formula contains
several special cases which have been discussed before in various
different contexts. These are an integral form of the Raychaudhuri
equation for null congruences, a special form of the constraints in
spherical symmetry and most notably the Bondi mass loss formula.  This
formula is obtained in the context of two-component spinors and the
formalism developed by Szabados \cite{Szabados-1994} on the geometry
of embedded two-surfaces. The main ingredient is the
Sparling-Nester-Witten identity. 

The paper is organised as follows: In section \ref{sec:prel} we
discuss the necessary background, in particular, the properties of
the two-dimensional spin connections known as the Sen connection and
the intrinsic spin connection. In section \ref{sec:intfor} we derive
the integral formula and in section \ref{sec:spec-case} we discuss
three special cases: the integrated Raychaudhuri equation, spherical
symmetry and null infinity. Throughout this work we use the
conventions of Penrose and Rindler \cite{PenroseRindler}.

\section{Geometric preliminaries}
\label{sec:prel}

Let $(M,g)$ be a Lorentz manifold with spin structure and let $\Sigma$
be an embedded hypersurface in $M$. We assume that $\Sigma$ is
diffeomorphic to $S\times I$ where $I$ is an open interval and $S$ is
a compact two-dimensional manifold so that $\partial \Sigma= S\times
\partial I$. Let $s$ be a parameter in $I$ and assume that the maps
$\Psi_s: S \to \Sigma$ provide the identification of $S$ with its
images $S_s$ embedded into $\Sigma$. We assume that all these images
are spacelike. Then each of these surfaces is the intersection of two
unique null hypersurfaces $\cN^+_s$, $\cN^-_s$, which exist in a
neighbourhood of $S_s$. At each point of $S_s$ there exist two unique
null directions which are orthogonal to the surface. The null
hypersurfaces are generated by the null geodesics which start at $S_s$
in these unique null directions. Note, that there is no preferred scaling
available on the generators. These null geodesic congruences are
(obviously) surface forming. 

We can use the null hypersurfaces to introduce coordinates in a
neighbourhood of $\Sigma$. Choose on $\Sigma$ a vector field $V$ with
$\langle ds, V\rangle=1$ and coordinates $(x^3, x^4)$ on $\Sigma$ in
such a way that $\langle dx^3, V\rangle=\langle dx^4, V\rangle=0$,
i.e., so that $V=\del_s$. Now we label each of the null hypersurfaces
$\cN^\pm_s$ by the value of $s$ on its intersection with
$\Sigma$. This defines two functions $u$ and $v$ in a neighbourhood of
$\Sigma$. The two families of null hypersurfaces are obtained by
$u=\mathrm{const}$ and $v=\mathrm{const}$, respectively, while
$\Sigma$ is defined by the equation $u=v$. Given a point $P$ near
$\Sigma$ we can now assign coordinates $(u,v,x^3,x^4)$ to it in the
following way: $P$ lies on exactly one null hypersurface from each
family. These define the values of $u$ and $v$. $(x^3,x^4)$ are the
coordinates of the intersection point with $\Sigma$ of the generator
through $P$ of the null hypersurface $u=\mathrm{const}$. The whole
purpose of this discussion was to show that we may assume that the
conormals of the null hypersurfaces are gradients. This will be of
importance later.

At each point $P$ of $S_s$ the tangent space is the direct sum of the
subspace which is tangent to $S_s$ at $P$ and its orthogonal
complement. Accordingly, the four-dimensional volume-element
$e_{abcd}$ can be split into two parts
\begin{equation}
  \label{2.1}
  e_{abcd} = 6 \,p_{[ab} m_{cd]},
\end{equation}
where $p_{ab}$ and $m_{ab}$ are two-forms. $m_{ab}$ is the induced
area-element on $S_s$ so it annihilates vectors orthogonal to $S_s$
and satisfies $m_{ab}m^{ab}=2$. Similarly, $p_{ab}$ vanishes when
contracted with vectors tangent to the surface and satisfies
$p_{ab}p^{ab}=-2$. Moreover, we have $m_{ab}p^{ac}=0$ and the fact
that $m_{ab} = -\frac12 e_{ab}{}^{cd} p_{cd}$ and $p_{ab} = \frac12
e_{ab}{}^{cd} m_{cd}$, so $m_{ab}$ and $p_{ab}$ are dual to each other.

In terms of spinors we have $p_{ab} = \frac12 \left(\gamma_{AB}
\eps_{A'B'} + \gamma_{A'B'} \eps_{AB} \right)$ with a symmetric spinor
$\gamma_{AB}$ which satisfies $\gamma_{AB} \gamma^{AC} =
-\eps_B{}^C$. From the duality we obtain $m_{ab} = i/2
\left(\gamma_{AB} \eps_{A'B'} - \gamma_{A'B'} \eps_{AB} \right)$. The
``surface spinor'' $\gamma_{AB}$ characterises the properties of the
surface in its four-dimensional surroundings in a way similar to the
normal vector of a hypersurface. In fact, the flag-pole of any
eigenspinor of $\gamma_{AB}$ (with eigenvalue $\pm1$) points into one
of the null directions defined by $S_s$.

We want to define two spin connections on the two-surface
$S_s$. First, there is the spin connection induced from the
four-dimensional spin connection, the so called two-dimensional Sen
connection $\nablab$ \cite{Szabados-1994}. Its action on spinor fields is
obtained from the four-dimensional spin connection simply by
restriction to the surface:
\begin{equation}
  \label{2.2}
  X^a\nablab_a \lambda_B = X^a\nabla_a \lambda_B \quad\mbox{for all
    $\lambda_B$ and all $X^a$ tangent to $S_s$.}
\end{equation}
It contains extrinsic information like the extrinsic curvatures and
the torsion of the normal bundle to $S_s$ in $M$. The other spin
connection is as intrinsically defined as possible by the requirements
that it agree with the Levi-Civita connection on $S_s$ when acting
on vectors which are tangent to $S_s$ and that it annihilate the
surface spinor $\gamma_{AB}$. These conditions do not completely
determine a unique spin connection. Two connections which satisfy the
above requirements differ by a term proportional to $\gamma_{AB}$,
i.e.,
\begin{equation}
  \label{2.3}
  \del_a \lambda_B - \del'_a \lambda_B = C_a \gamma_{BC} \lambda^C.
\end{equation}
The one-form $C_a$ cannot be fixed by intrinsic properties. However,
we can define a spin connection which satisfies these requirements
from the two-dimensional Sen connection. First we define the spinor
valued one-form $Q_{aB}{}^C = \frac12 \gamma_A{}^C \nablab_a
\gamma_B{}^A$ and then the connection $\del$ by
\begin{equation}
  \label{2.4}
  \del_a \lambda_B = \nablab_a \lambda_B + Q_{aB}{}^C \lambda_C.
\end{equation}
The spin connection thus defined annihilates $\gamma_{AB}$ and agrees
with the intrinsic Levi-Civita connection when acting on tangent
vectors of $S_s$. However, it still contains some extrinsic piece
which shows up in its curvature
\begin{equation}
  \label{2.5}
  \del_a \del_b \lambda_C - \del_b \del_a \lambda_C = -S_{abC}{}^E
  \lambda_E = (\del^a C_a + \frac{i}{4} \cR) m_{ab} \gamma_C{}^E \lambda_E.
\end{equation}
Here the function $\cR$ is the scalar curvature of $S_s$ and $C_a$ is
a local representation of a $SO(1,1)$ connection in the normal bundle
of $S_s$. This ambiguity does not affect the ensuing discussion, so we
will not try to fix $C_a$ any further.

For later use we introduce a spin-frame $(o^A,\iota^A)$ adapted to the
surface $S_s$ by writing $\gamma_A{}^B = o_A\iota^B + \iota_A
o^B$. Thus, $o^A$ and $\iota^A$ are eigenspinors of $\gamma_A{}^B$
with eigenvalues $-1$ and $+1$, respectively. Let $l^a$, $n^a$, $m^a$
and $\mb^a$ be the corresponding null tetrad. Then $m^a$ and $\mb^a$
are tangent to $S_s$ while $l^a$ and $n^a$ point along the generators
of the null hypersurfaces $\cN^\pm_s$. The spin-frame is defined away
from the surface by taking the spinors to be parallel along their
respective flag-poles, i.e., we impose the propagation equations $Do^A
=0$ and $D'\iota^A=0$. This definition yields a spin-frame in a
neighbourhood of $\Sigma$ which is fixed up to the scalings
\begin{equation}
  \label{2.6}
  o^A\mapsto c o^A,\quad \iota^A \mapsto \frac1c \iota_A,
\end{equation}
where $c$ is any function on $\Sigma$. This is the natural setup for
the GHP-formalism \cite{PenroseRindler,GHP-1973} which will be used to some
extent in the following discussion.  With $e_{abcd} = -24il_{[a}n_b
m_c \mb_{d]}$ the induced area-elements are $m_{ab} =
-2im_{[a}\mb_{b]}$ and $p_{ab} = 2l_{[a} n_{b]}$.

In this spin-frame the spinor-valued one-form $Q_{aB}{}^C$ can be
determined from the four-dimensional spin connection, with $\rho$ and
$\sigma$ being the usual spin-coefficients
\begin{equation}
  \label{2.7}
  Q_{aB}{}^C = - \left( m_a \rho + \mb_a \sigma \right) \iota_B  \iota^C
               + \left( m_a \sigma' + \mb_a \rho' \right) o_B  o^C.
\end{equation}
The action of the intrinsic connection on the spin-frame can be found
from the fact that it annihilates the  surface spinor
$\gamma_A{}^B$. For any eigenspinor $\lambda_B$ of $\gamma_A{}^B$ we
have 
\begin{equation}
  \label{2.8}
  \gamma_A{}^B \lambda_B = \pm \lambda_A \Rightarrow
  \gamma_A{}^B \del\lambda_B = \pm \del\lambda_A \Rightarrow
  \del\lambda_A = \Gamma \lambda_A
\end{equation}
for some one-form $\Gamma$. Therefore, we obtain
\begin{align}
  \label{2.9}
  \del o_A &= \Gamma o_A, \nonumber\\
  \del \iota_A &= -\Gamma \iota_A.
\end{align}
The one-form $\Gamma$ can be expanded in terms of the complex basis
$(m_a, \mb_a)$ with the coefficients given by the usual
spin-coefficients: $\Gamma_a = -m_a \alpha - \mb_a \beta$. This
one-form represents both the extrinsic $SO(1,1)$ connection and the
intrinsic $SO(2)$ connection which are combined in $\Gamma$ as (twice)
its real and imaginary parts respectively.

For any spinor field $\lambda_A = \lambda_1 o_A - \lambda_0 \iota_A$
we have the formula
\begin{equation}
  \label{2.9-1}
  \del_a\lambda_A = -\left( m_a \eth' \lambda_1 + \mb_a \eth \lambda_1
    \right) o_A + \left( m_a \eth' \lambda_0 + \mb_a \eth \lambda_0
    \right) \iota_A,
\end{equation}
and similar expressions for spinor fields with other valences.

We end this section with the Sparling-Nester-Witten identity. Given
two spinor fields $\lambda_A$ and $\mu_{A'}$, we can
define a two-form by $L=-i\mu_{A'} \dd \lambda_A \theta^{AA'}$. The
$\dd$ here is the covariant exterior derivative of the
four-dimensional spin connection acting on spinor valued differential
forms, the $\theta^{AA'}$ are essentially the van-der-Waerden symbols
and the product between differential forms is understood to be the
Grassmann product. Computing the exterior derivative of $L$ we find
\begin{equation}
  \label{2.10}
  d\,L = \dd \,L = -i\dd \mu_{A'} \dd \lambda_A \theta^{AA'} 
  - i \mu_{A'} \dd^2 \lambda_A \theta^{AA'} = S + E.
\end{equation}
The three-form $E$ contains only curvature terms and, remarkably,
only in the form of the Einstein tensor. In fact, one can show that
(e.g., \cite{PenroseRindler}) 
$E=-\frac12 V^aG_a{}^b \Sigma_b$ with $V^a = \lambda^A \mu^{A'}$ and
$\Sigma_a = \frac16 e_{abcd} \theta^b \theta^c \theta^d$. Using the
Einstein equations in the form $G_{ab}=-8\pi G T_{ab}$ the identity
becomes
\begin{equation}
  \label{2.11}
  d\,L = -i\dd \mu_{A'} \dd \lambda_A \theta^{AA'} + 4\pi G\,
  V^aT_a{}^b \Sigma_b. 
\end{equation}
This identity is the basis for the spinorial proofs of positivity of
mass in General Relativity which are obtained by choosing the spinor
fields appropriately.

\section{The integral formula}
\label{sec:intfor}

Consider the integral defined by
\begin{equation}
  \label{3.1}
  I(s) = \oint_{S_s}\, L.
\end{equation}
We are interested in the change of $I$ with the parameter $s$. To
compute it we choose a vector field $Z^a$ on $\Sigma$ so that $Z^a
\nabla_a s = 1$. Such vector fields exist because $\frac{\del}{\del
  s}$ is such a vector field but we could just as well choose any
other field which differs from it by terms tangent to the
foliation. Now recall that for any two-form $\omega$ we have
\begin{equation}
\label{3.2}
  \frac{d}{ds}\oint \omega = \oint L_Z \omega = \oint i_Z d\omega.
\end{equation}
Here we have used the formula that the Lie derivative of a
differential form is $L_Z\omega = d i_Z \omega + i_Z d\omega$ and the
fact that the two-surfaces are closed. This formula also shows that we
can assume without loss of generality that the vector field $Z^a$ is
orthogonal to the foliation: components of $Z^a$ tangent to the
two-surfaces $S_s$ would correspond to applying an infinitesimal
diffeomorphism to the integrand which does not change the value of the
integral because there are no boundaries. Now the identity
\eqref{2.10} implies
\begin{equation}
  \label{3.3}
\frac{d}{ds}\oint L = \oint i_ZdL =   \oint i_Z S + \oint i_Z E.
\end{equation}
The last term in this formula is easily evaluated. Note that $i_Z
\theta^a = Z^a$ and $i_Z \Sigma_a = \frac12 e_{abcd} Z^b \theta^c
\theta^d$ which, after restriction to the two-surface, becomes
\begin{equation}
  \label{3.3-1}
  3p_{[ab}m_{cd]} Z^b \theta^c \theta^d = p_{ab} Z^b d^2A.
\end{equation}
Thus,
\begin{equation}
  \label{3.4}
  \oint i_Z E = 4\pi G \oint \left( T_a{}^b V^a p_{bc} Z^c\right) d^2A.
\end{equation}
The evaluation of the first term in \eqref{3.3} is more
complicated. We have
\begin{align}
  \label{3.5}
  i_Z S &= i_Z\left(-i \dd\mu_{A'}\,\dd\lambda_A\,\theta^{AA'} \right) 
  \nonumber\\
        &= -i Z^e\nabla_e \mu_{A'} \dd\lambda_A\theta^{AA'} +
          i Z^e\nabla_e \lambda_A \dd\mu_{A'}\theta^{AA'} -
          i \dd\mu_{A'}\,\dd\lambda_A\,Z^{AA'}
\end{align}
After restriction to the surface the $\dd$ becomes the exterior
covariant differential with respect to the two-dimensional Sen
connection and we can express it in terms of the intrinsic spin
connection using the relation $\dd\lambda=\del \lambda_A -
Q_A{}^C\lambda_C$ where we denote the exterior covariant differential
with respect to the intrinsic spin connection by $\del$ and where
$Q_A{}^B = \theta^a Q_{aA}{}^B$. Then the last term in \eqref{3.5}
becomes
\begin{multline}
  \label{3.6}
  \dd\mu_{A'}\,\dd\lambda_A\,Z^{AA'} 
  = \Big(\del \mu_{A'} - \overline{Q}_{A'}{}^{C'} \mu_{C'}\Big)
  \Big( \del \lambda_A - Q_A{}^C \lambda_C \Big) Z^{AA'} \\
  = \del \mu_{A'} \del \lambda_A  Z^{AA'} 
   - \del \mu_{A'} Q_A{}^C \lambda_C Z^{AA'} \\
  + \del \lambda_A  \overline{Q}_{A'}{}^{C'} \mu_{C'} Z^{AA'}
   - Q_A{}^C \lambda_C \overline{Q}_{A'}{}^{C'} \mu_{C'} Z^{AA'}.
\end{multline}
The first term above can be rewritten as follows
\begin{multline}
  \label{3.7}
  \del \mu_{A'} \del \lambda_A  Z^{AA'} 
  = \frac12\,d\left( \mu_{A'} \del \lambda_A Z^{AA'} 
    - \lambda_A \del \mu_{A'} Z^{AA'} \right) \\
  + \frac12 \left\{ \lambda_A \del^2 \mu_{A'} Z^{AA'}
    - \mu_{A'} \del^2 \lambda_A Z^{AA'} 
    - \lambda_A \del \mu_{A'} \del Z^{AA'}
    + \mu_{A'} \del \lambda_A \del Z^{AA'} \right\}.
\end{multline}
On the right hand side of this equation, the total derivative term can
be dropped because it does not contribute to the integral. Hence we
get after collecting similar terms
\begin{multline}
\label{3.8}
  i_Z S = -i\left\{ - Z^e\nabla_e \mu_{A'} Q_A{}^E\lambda_E\theta^{AA'}   
      + Z^e\nabla_e \lambda_A \overline{Q}_{A'}{}^{E'} \mu_{E'}
      \theta^{AA'} \right\} \\
    -i \left\{ \overline{Q}_{A'}{}^{E'}  Q_A{}^E \mu_{E'} \lambda_E
      Z^{AA'} + \frac12  \lambda_A \del^2 \mu_{A'} Z^{AA'}
      - \frac12 \mu_{A'} \del^2 \lambda_A Z^{AA'} \right\} \\
    -i \left\{ \frac12 \mu_{A'} \del \lambda_A \del Z^{AA'} 
      - \overline{Q}_{A'}{}^{E'}  \mu_{E'} \del \lambda_A Z^{AA'}
      + Z^e \nabla_e \mu_{A'} \del \lambda_A \theta^{AA'} \right\} \\
    -i \left\{-\frac12 \lambda_A \del \mu_{A'} \del Z^{AA'}
      +  Q_A{}^E \lambda_E \del \mu_{A'} Z^{AA'}
      - Z^e \nabla_e \lambda_A \del \mu_{A'} \theta^{AA'} \right\}.
\end{multline}
This is the general form of this term. Further simplifications can
only be obtained by specialising the spinor fields $\mu_{A'}$ and
$\lambda_A$. Usually this is done by imposing the Witten equation on
the spinor fields. Here, we will proceed somewhat differently. The
discussion is simplified by the following trivial observations:
\begin{itemize}
\item $\del$ leaves the eigenspaces of $\gamma_A{}^B$ invariant,
\item $\del$ leaves the splitting of the tangent space at points of
  $S_s$ into vertical and horizontal parts invariant,
\item $Q_A{}^B$ interchanges the eigenspaces of $\gamma_A{}^B$,
\item $V^a = \lambda^A \mu^{A'}$ is horizontal/vertical iff
  $\lambda^A$ and $\mu^{A'}$ are in different/same eigenspace(s).
\end{itemize}
Let us now assume that the spinor fields are both in the eigenspace of
$\gamma_A{}^B$ spanned by $o_A$, hence $\lambda_A=\lambda o_A$ and
$\mu_{A'}=\mub o_{A'}$. In this case, the flag-poles of both
$\lambda_A$ and $\mu_{A'}$ point along the generators of the same
outgoing null hypersurface $\cN^+$. From the properties of the
intrinsic derivative we find
\begin{equation}
\label{3.9}
\overline{Q}_{A'}{}^{E'} \mu_{E'} \del \lambda_A Z^{AA'}
=0=
Q_A{}^E \lambda_E \del \mu_{A'} Z^{AA'}.
\end{equation} 
We further assume that $Z^a$ is orthogonal to the surfaces $S_s$ so
that we can write it as $Z^a = Zl^a + Z'n^a$. Using \eqref{2.9-1} and
similar expansions for $\mu_{A'}$ and $Z^{AA'}$ we obtain
\begin{multline}
\label{3.9-1}
i_ZS = 
  \lambda \rho \left( \iota^{A'} Z^e \nabla_e \mu_{A'} \right) \,d^2A 
  + \mub \rho \left( \iota^A Z^e \nabla_e \lambda_A  \right) \, d^2A \\
+ \frac18 Z' \mub \lambda \cR\, d^2A  
  + Z \mub \lambda \left( \sigma\sigmab - \rho^2 \right)\, d^2A \\
  + \frac12 \left( \mub \eth' \lambda \eth Z' 
    - \mub \eth \lambda \eth' Z' 
    - \lambda \eth' \mub \eth Z' 
    + \lambda \eth \mub \eth' Z'
    - 2 \taub \mub \eth \lambda Z'
    - 2 \tau \lambda \eth' \mub Z' \right)\,d^2A.
\end{multline}

In this formula the last term can be simplified considerably by
appropriate choice of the scalings of the spinor fields. First, we
note that we can choose them completely freely on each surface
$S_s$. Furthermore, we can exploit the fact that the vector field
$Z_a$ is hypersurface orthogonal. This leads to the equations
\begin{equation}
  \label{3.10}
  \eth Z + \taub' Z = 0,\qquad  \eth Z' + \tau Z' = 0.
\end{equation}
Inserting these into the last term in \eqref{3.9-1} yields the
expression
\begin{equation}
  \label{3.11}
   -\frac12 Z' \mub \lambda \left\{
     \tau \left( \frac{\eth'\lambda}{\lambda}
    + \frac{\eth' \mub}{\mub} \right)
    + \taub \left( \frac{\eth \lambda}{\lambda}
    + \frac{\eth \mub}{\mub} \right)
 \right\}\,d^2A 
\end{equation}
for that term. As pointed out in the previous section, we can assume
without loss of generality that the conormal of $\cN^-$ is a
gradient. If $\lambda_A$ is scaled so that its flag-pole coincides on
$\Sigma$ with that gradient then the following equations must hold
there:
\begin{align}
  \label{3.12}
  \eth \left(\lambda\lambdab\right) = \tau\lambda\lambdab, \\
  \label{3.13}
  \thorn \left(\lambda\lambdab\right) = 0.
\end{align}
Obviously, these are conditions only for the extent $\lambda\lambdab$
of the flag-pole of the spinor field. The first equation above tells
us how to fix it on $S_s$ while the second equation specifies how to
propagate it along the generators of $\cN^-$. Note, that if we choose
$\mu_{A'} = \lambdab_{A'}$ on $\Sigma$ then \eqref{3.11} simplifies
further by means of \eqref{3.12} to yield the final expression
\begin{equation}
  \label{3.14}
  - Z' \lambda \lambdab \left(\tau\taub\right) \,d^2A. 
\end{equation}
Henceforth, we will assume that the spinor fields have been chosen in
that way on $\Sigma$. Therefore, \eqref{3.12} holds on $\Sigma$. We
will not impose \eqref{3.13} because it is not needed. Note further,
that there is some freedom left in choosing $\lambda$. First of all,
it can be multiplied by any phase function on $\Sigma$ and, what is
more important, it is defined only up to the multiplication with any
function $f(s)$ on $\Sigma$ which is constant on the surfaces $S_s$.

Let us now consider the first two terms in \eqref{3.9-1}. These can
expanded to yield the expression
\begin{equation}
  \label{3.14-1}
  \rho  \left( 
    Z^e \nabla_e \left( \lambda \lambdab \right)
    + Z \lambda \lambdab (\eps + \epsb) 
    + Z' \lambda \lambdab(\gamma + \gammab)  \right) \, d^2A,
\end{equation}
where we have introduced the spin-coefficients $\eps$ and
$\gamma$. However, from the definition of the spin-frame these
coefficients are seen to vanish because on $\Sigma$ we have $Do^A=0$
and $D'\iota^A=0$.

This concludes the evaluation of the right hand side of \eqref{3.3} in
the case where both spinor fields are chosen to be tangent to the same
null hypersurface. In this case, we obtain for the left hand side of
\eqref{3.3}
\begin{equation}
  \label{3.15}
  -i \frac{d}{d s} \oint \mu_{A'} \dd \lambda_A \theta^{AA'} =
  \frac{d}{d s} \oint \lambda \lambdab \rho \,d^2A.
\end{equation}

So we can finally put everything together to obtain the integral
formula
\begin{multline}
  \label{3.16}
  \frac{d}{d s} \oint \phi \rho \,d^2A = 
  \oint \rho \dot\phi \,d^2A
  + Z' \phi \oint \left( \frac\cR8   
    - \tau\taub \right)\,d^2A \\
  + \oint Z \phi \left( \sigma\sigmab - \rho^2 \right)\, d^2A
  + 4\pi G \oint \phi \left( l^a T_a{}^b p_{bc} Z^c\right) d^2A,
\end{multline}
where $\phi$ is a positive $(-1,-1)$-function on $\Sigma$ which
satisfies the equation $\eth \phi = \tau \phi$ on each two-surface
$S_s$ and $\dot\phi=Z^e\nabla_e \phi$. Note, that this equation always
has solutions because it corresponds to the geometric statement that
one can always choose the tangents of the outgoing null geodesics on
$S_s$ to coincide with the gradients of the null hypersurface which is
generated by them. These tangent vectors are given by $\phi l^a$. We
have pulled out the factor $Z' \phi$ from the integral because, as a
consequence of \eqref{3.10} and \eqref{3.12}, we have $Z'\phi =
\mathrm{const.}$ on the surfaces $S_s$. This equality corresponds to
the fact that the function which defines the null hypersurface
foliation can be chosen to agree with $s$ on $\Sigma$ (cf. the
discussion at the beginning of section \ref{sec:prel}). Note, that we
have not yet specified the scaling of the spin-frame. Under special
circumstances this can be chosen to further simplify the
formula. Furthermore, the term involving the scalar curvature
integrates to a constant by the Gau\ss-Bonnet theorem.

The quantity on the left hand side is the integrated divergence of
that null geodesic congruence emanating from $S_s$ for which the
tangent vectors coincide with the gradient of the null hypersurface
that is formed by the congruence. On the right hand side we find terms
which have to do with the geometry of the two-surfaces, namely the
scalar curvature $\cR$ and the determinant $\rho^2-\sigma\sigmab$ of
the extrinsic curvature with respect to the corresponding null
normal. Note, that this expression occurs in the Gau\ss\ maps defined
in \cite{Tod-1992} and has to do with the convexity of the
two-surfaces. The $\tau$-terms and the term involving the derivative
of $\phi$ have no immediate geometric meaning. However, it is worth to
point out that on a spacelike hypersurface $\Sigma$ under the
assumption of the dominant energy condition the $\tau$-term combines
with the energy-momentum term with the same sign. So it might be
possible to view that term as some kind of gravitational contribution
to the total energy. The term involving the derivative of $\phi$ is
present because it guarantees the invariance under reparametrisation of
the foliation.

In the case where the spinor fields are aligned along the other null
hypersurface $\cN^+$ (so that $\lambda_A = \mub_{A} = \mub \iota_A$)
we use the analogous assumptions to arrive at the formula
\begin{multline}
  \label{3.17}
  \frac{d}{d s} \oint \phi' \rho' \,d^2A = 
  \oint \rho' \dot\phi' \,d^2A
  +  Z \phi' \oint  \left( \frac\cR8   
    - \tau'\taub' \right)\,d^2A \\
  + \oint Z' \phi' \left( \sigma'\sigmab' - {\rho'}^2 \right)\, d^2A
  - 4\pi G \oint \phi' \left( n^a T_a{}^b p_{bc} Z^c\right) d^2A,
\end{multline}
where now $\phi'$ is a $(1,1)$-function satisfying $\eth' \phi' = \tau'
\phi'$ on each two-surface $S_s$.

\section{Special cases}
\label{sec:spec-case}

\subsection{On a null hypersurface}
\label{sec:nullhsf}

First, we look at the case where $\Sigma$ is itself a null
hypersurface. Let $l_a$ be the normal to $\Sigma$ scaled so that it is
a gradient. Then $l^a$ is tangent to the null geodesics generating
$\Sigma$ and we choose an affine parameter $s$ along each
generator. Fix a two-dimensional surface $S$ orthogonal to each
generator and define a foliation of $\Sigma$ by the surfaces $S_s$ of
constant affine parameter $s$ with $S=S_0$. Thus we can take
$\phi=1$ and $Z^a=l^a$, i.e., $Z'=0$ and $Z=1$. Then \eqref{3.16}
results in the formula
\begin{equation}
  \label{4.1}
  \frac{d}{d s} \oint \rho \,d^2A = 
   \oint \left( \sigma\sigmab - \rho^2 \right)\, d^2A
  + 4\pi G \oint  T_{ab} l^a l^b \, d^2A.  
\end{equation}
This formula is one of the optical equations \cite{Sachs-1961}
integrated over $S_s$. It is the equation which governs the focusing
of a bundle of light rays. This can be seen by noting that the left
hand side can be rewritten as
\begin{equation}
  \label{4.2}
  \frac{d}{d s} \oint \rho \,d^2A = \oint \left( D\rho - 2 \rho^2
  \right)\,d^2A.  
\end{equation}
by an argument similar to equation \eqref{3.2}. Thus we obtain 
\begin{equation}
  \label{4.3}
\oint D\rho \,d^2A  = \oint \left( \rho^2 + \sigma\sigmab \right)\, d^2A
  + 4\pi G \oint  T_{ab} l^a l^b \, d^2A.
\end{equation}

The formula \eqref{3.17} can be specialised in this case to yield
\begin{equation}
  \label{4.4}
  \frac{d}{d s} \oint \rho' \,d^2A = 
  \oint \left(\frac18 \cR - \tau'\taub'\right) \,d^2A 
  - 4\pi G \oint T_{ab} n^a l^b d^2A.
\end{equation}
Here, we have used the fact that $\phi'$ is constant on the
two-surfaces because $Z=1$.

\subsection{Spherical symmetry}
\label{sec:sphersymm}

The next special case occurs in space-times with spherical
symmetry. We take $\Sigma$ as a spacelike hypersurface which is
foliated by the spheres $S_s$ of symmetry. Due to the symmetry all
quantities with a non-vanishing spin-weight are zero and, in addition,
all integrands are constant on $S_s$ so that the integrals yield the
integrand times the area $A(s)=4\pi R(s)^2$ of the surface.
Then \eqref{3.16} gives 
\begin{multline}
  \label{4.5}
  \frac{d}{d s} \left( \phi \rho R^2 \right) =  
  \left( \rho \dot\phi + \frac18 \cR \phi Z' \right)R^2
  -  Z \phi (\rho R)^2
  + 4\pi G  \phi \left( l^a T_a{}^b p_{bc} Z^c\right) R^2.
\end{multline}
Since we are free to multiply $\phi$ by any function of $s$, we may
take $\phi=1/R$. Next, we take the parameter $s$ to be the proper
length along a radial geodesic so that $Z^a$ is a unit vector
field. Furthermore, we choose the spin-frame so that $Z'=-Z=1/\sqrt2$,
thus breaking the scaling invariance \eqref{2.6}. Then we also have
$p_{bc}Z^c = Zl_b - Z'n_b = -Z'(l_b + n_b) = -t^b$ where $t_b$ is the
future pointing timelike unit normal vector to $\Sigma$. Finally,
inserting the scalar curvature $\cR=4/R^2$ of the surface we obtain
\begin{equation}
  \label{4.6}
  \frac{d}{d s} \left( \rho R \right) = 
  \frac1{\sqrt2R}\left( \frac12 + (\rho R) (\rho' R)  
  \right)  - 4\pi G  R T_{ab} l^a t^b.
\end{equation}
Here, we have used the formula $Z^e\nabla_e R=(\rho - \rho')
R/\sqrt2$, a consequence of the formula for the rate of change in the
area element along a vector field \cite{PenroseRindler}. Let us now
define the quantities $\omega_+ = 2\sqrt2\rho R$ and $\omega_- =
-2\sqrt2\rho' R$ then we can establish complete agreement with
\cite{OMurchadhaMalec-1994} where an alternative form of the constraint
equations for spherically symmetric space-times has been
derived. These equations are
\begin{align}
  \label{4.7}
  \frac{d\omega_+}{d s}  &= 
  \frac1{4R}\left( 4 - \omega_+ \omega_-
  \right)  - 8\pi G  R T_{ab} (\sqrt2l^a) t^b, \\
  \frac{d\omega_-}{d s} &= 
  \frac1{4R}\left( 4 - \omega_+ \omega_-
  \right)  - 8\pi G  R T_{ab} (\sqrt2n^a) t^b.
\end{align}
These two equations have several important consequences. First,
assuming that $\Sigma$ is topologically $\RR^3$, asymptotically flat
and regularity at $R=0$ of $\Sigma$ and the dominant energy condition
one can deduce global bounds for the quantities $\omega_\pm$. These in
turn imply statements about the occurrence of trapped surfaces and
singularities in the domain of dependence of $\Sigma$.

Furthermore, one can easily prove the Penrose inequality in this case
\cite{Niall}. Suppose that $\Sigma$ contains an apparent horizon and
is asymptotically flat. Consider the Hawking mass for a sphere $S_s$
which is given in the present context by
\begin{equation}
  \label{4.8}
  m_H(s) = \frac{R}{8G}\left(4 - \omega_+ \omega_-\right).
\end{equation}
Its derivative along $Z^a$ is
\begin{equation}
  \label{4.9}
  \frac{d m_H}{d s} = \pi R^2 T_{ab} l^b \left(l^a \omega_-  +
    n^a \omega_+ \right).
\end{equation}
At the outermost apparent horizon with area $A_0=4\pi R_0^2$ the mass
is $m_H=R_0/(2G)$ while at spatial infinity we have $m_H=m_{ADM}$. In
the region in between both $\omega_\pm$ are positive. From \eqref{4.9}
we find that
\begin{equation}
  \label{4.10}
  \frac{d m_H}{d R} = 4 \pi R^2 T_{ab} l^b H^a \ge 0.
\end{equation}
The positivity follows from the fact that under the conditions stated
above
\begin{equation}
  \label{4.11}
  H^a = \frac{\omega_-}{\omega_+ + \omega_-} l^a +   
        \frac{\omega_+}{\omega_+ + \omega_-} n^a
\end{equation}
is a future pointing timelike vector so that the sign is a consequence
of the dominant energy condition. From this we obtain the inequality
$R_0 \le 2Gm_{ADM}$ or
\begin{equation}
  \label{4.12}
  A_0 \le 16\pi G^2 m_{ADM}^2.
\end{equation}

\subsection{At null infinity}
\label{sec:scri}

Our last specialisation occurs in asymptotically flat space-times which
admit a regular null infinity $\scri$ when $\Sigma$ coincides with
$\scri$. To treat this case we proceed as follows. We introduce Bondi
coordinates $(u,r,\zeta,\zetab)$ and the usual Bondi null tetrad in a
neighbourhood of $\scri$. We evaluate the integral formula on the
hypersurfaces $\Sigma_r$ of constant $r$ and then take the limit
$r\to\infty$. The evaluation of the formulae has to be done only to
leading order in $1/r$. The asymptotic solution of the vacuum
equations is taken from \cite{NewmanTod-1980}. 

Each hypersurface $\Sigma_r$ is foliated by its intersections $S_u$
with the null hypersurfaces $u=\mathrm{const}$ so we take $s=u$. Since
the Bondi tetrad is not yet adapted to these surfaces we need to
perform a null rotation around $l^a$ to put $n^a$ orthogonal to
$S_u$. Then the covariant form of the tangent vectors to the outgoing
null hypersurfaces is
\begin{equation}
  \label{4.13}
  l = du, \qquad n = dr - (U - \omega \omegab) du.
\end{equation}
The spin coefficients need to be changed accordingly. In particular,
we get to leading order
\begin{align}
  \label{4.14}
  \rho &= -r^{-1} + O(r^{-3}), \\
  \rho' &= \frac12r^{-1} 
  + \left(\Psi_2^0 + \sigma^0 \dot{\sigmab}^0 + \eth^2 \sigmab^0
  \right)r^{-2} + O(r^{-3}), \\
  \sigma &= \sigma^0r^{-2} + O(r^{-3}),\\
  \sigma' &= - \dot{\sigmab}^0 r^{-1} + O(r^{-2}), \\ 
  \tau' &= -\eth' \sigma^0 r^{-2} + O(r^{-3}).
\end{align}
Here, $\eth$ is defined with respect to the unit sphere. Note, that
$\rho'$ is real due to the null rotation and by virtue of an equation
on $\scri$ which follows from the asymptotic solution of the vacuum
field equations.

The vector field $Z^a$ is given by $Z^a = n^a - (U -
\omega\omegab) l^a$ and the function $\phi'$ on $\Sigma_r$ can be
taken to have the form $\phi' = \phi_0(u) + \phi_1(u) r^{-1} +
O(r^{-2})$, where $\eth' \phi_1 = (\eth' \sigma^0) \phi_0$. Finally,
the scalar curvature of the two-surfaces $S_u$ is $\cR=4r^{-2} +
O(r^{-3})$. Inserting these expansions into \eqref{3.17} gives
\begin{multline}
  \label{4.15}
  \frac{d}{d u} \oint \left\{ \frac12 \left(r\phi_0 + \phi_1\right) +
    \left(\Psi_2^0 + \sigma^0 \dot{\sigmab}^0 + \eth^2
      \sigmab^0 \right) \phi_0 
    \right\}\,d^2\Omega = \\ 
    \oint \frac12 \left( r\dot\phi_0 + \dot\phi_1 \right) 
    + \left(\Psi_2^0 + \sigma^0 \dot{\sigmab}^0 + \eth^2 \sigmab^0\right)
    \dot\phi_0\,d^2\Omega\\
    + \oint \frac14 \phi_0 \,d^2\Omega
    + \oint \phi_0 \left(\dot{\sigmab}^0\dot{\sigma}^0 - \frac14
    \right)\,d^2\Omega  + O(r^{-1}).
\end{multline}
Here $d^2\Omega$ is the surface element of the unit sphere.  This
equation can be simplified. Note, that the $O(r)$-terms cancel and
that the $\eth^2\sigmab^0$ term vanishes upon integration over $S_u$
so that in the limit $r\to \infty$ we obtain the Bondi mass loss
formula
\begin{equation}
  \label{4.16}
  \frac{d}{du} m_B = \frac{d}{d u} \left\{-\frac{1}{4\pi G} \oint
    \Psi_2^0 + \sigma^0 \dot{\sigmab}^0  \,d^2\Omega \right\}
  = - \frac{1}{4\pi G} \oint \dot{\sigmab}^0\dot{\sigma}^0
  \,d^2\Omega.  
\end{equation}
The other integral formula \eqref{3.17} contains no information when
$\Sigma$ coincides with $\scri$.

It is worthwhile to point out that in the limit when $\Sigma$ coincides
with $\scri$ it is a null hypersurface and that the integral formula
is then exactly the one discussed as the first special case, namely the
integrated focusing equation. This seems to suggest that the
gravitational energy flux and focusing power are really closely
related. This has been discussed before \cite{Penrose-1966} and there
have also been attempts to prove positivity of mass from the focusing
properties of light rays \cite{PenroseSorkinWoolgar}.

\section{Discussion}
\label{sec:disc}

We have derived an integral formula valid on an arbitrary
hypersurface $\Sigma$ in a space-time which satisfies Einstein's
equation. It is  obtained in a straightforward way from the
Sparling-Nester-Witten identity by a particular choice of the spinor
fields. In contrast to the usual procedure of fixing the spinor fields
by having them obey a differential equation here they are chosen in a
geometric way. It is not clear whether this fixing is optimal. It
might well be that there is a different choice which eliminates the
somewhat arbitrary and uncontrollable functions $\phi$ and $\phi'$. 
Furthermore, in the final form there is no essentially spinorial
feature left in the integral formula. It might be worthwhile to see
whether one can derive the formula without using spinors.

Of course, there are many questions left. We have seen that the
integral formula reduces to the special form of the constraint
equations derived by \cite{OMurchadhaMalec-1994} in the case of spherical
symmetry with spacelike $\Sigma$. They can derive bounds for the
optical scalars $\omega_\pm$ when $\Sigma$ is a regular embedded
hypersurface which allows them to obtain statements about the
occurrence of trapped surfaces and singularities in the domain of
dependence of $\Sigma$. Is it possible to get bounds on the integral
expressions or some variant of those, too? Can one prove the Penrose
inequality using these integral formulae in a way similar to the one
given above?

We have not looked at the other possibilities to choose the spinor
fields. The cases where the two spinor fields are aligned along
different null hypersurfaces do not seem to lead to any similarly nice
integral formula.

\section*{Acknowledgements}
This work was started during a stay at the Erwin Schr\"odinger
Institute, Vienna. I am grateful to the organisers of the two
workshops, H. Urbantke and R. Beig, for providing the stimulating
atmosphere. I also wish to thank K.~P.~Tod, E.~T.~Newman, R.~Penrose
and in particular L.~Szabados for discussions and for reading a
preliminary version of this article.


\begin{thebibliography}{1}

\bibitem{Szabados-1994}
L.~B. Szabados.
\newblock {\sl Two-dimensional Sen connections in General Relativity}.
\newblock {\em Class. Quant. Grav.}, {\bf 11}, p.~1833--1846, 1994.

\bibitem{PenroseRindler}
R.~Penrose and W.~Rindler.
\newblock {\em Spinors and Spacetime}, volume 1,2.
\newblock Cambridge University Press, 1984, 1986.

\bibitem{GHP-1973}
R.~P. Geroch, A.~Held, and R.~Penrose.
\newblock {\sl A space-time calculus based on pairs of null directions}.
\newblock {\em J. Math. Phys.}, {\bf 14}, p.~874--881, 1973.

\bibitem{Tod-1992}
K.~P. Tod.
\newblock {\sl The {S}t{\"u}tzfunktion and the cut function}.
\newblock In A.~I. Janis and J.~R. Porter, editors, {\em Recent advances in
  General Relativity}. Birkh{\"a}user, Boston, 1992.

\bibitem{Sachs-1961}
R.~K. Sachs.
\newblock {\sl Gravitational waves in general relativity. {VI} {T}he outgoing
  radiation condition}.
\newblock {\em Proc. Roy. Soc. London~A}, {\bf 264}, p.~309--338, 1961.

\bibitem{OMurchadhaMalec-1994}
E.~Malec and N.~{\'O}. Murchadha.
\newblock {\sl Optical scalars and singularity avoidance in spherical
  spacetimes}.
\newblock {\em Phys. Rev.~D}, {\bf 50}, p.~6033--6036, 1994.

\bibitem{Niall}
N.~{\'O}. Murchadha.
\newblock {\em private communication}, 1997.

\bibitem{NewmanTod-1980}
E.~T. Newman and K.~P. Tod.
\newblock {\sl Asymptotically flat spacetimes}.
\newblock In A.~Held, editor, {\em General Relativity and Gravitation. One
  hundred years after the birth of Albert Einstein}. Plenum, New York, 1980.

\bibitem{Penrose-1966}
R.~Penrose.
\newblock {\sl General-Relativistic energy flux and elementary optics}.
\newblock In B.~Hoffmann, editor, {\em Perspectives in Geometry and
  Relativity}, pages 259--274. Indiana University Press, 1966.

\bibitem{PenroseSorkinWoolgar}
R.~Penrose, R.~D. Sorkin, and E.~Woolgar.
\newblock {\sl A positive mass theorem based on the focusing and retardation of
  null geodesics}.
\newblock preprint.

\end{thebibliography}
\end{document}